\documentclass[runningheads]{llncs}
\usepackage[T1]{fontenc}
\usepackage{multirow}
\usepackage{amsmath}
\usepackage{url}
\usepackage{hyperref}
\usepackage{graphicx}
\begin{document}
\title{Tuning Quantum MPS}
%
%

 \author{Anna Leonteva\inst{1}\orcidID{0000-0001-9521-0630} \and
 Maxime Outteryck \inst{1}\orcidID{0009-0002-7428-1776} \and
 Guido Masella\inst{1}\orcidID{0000-0001-7108-2996}}

\authorrunning{A. Leonteva et al.}

\institute{QPerfect, 23 Rue du Loess, 67000 Strasbourg, France
\email{contact@qperfect.io}\\
\url{https://qperfect.io/}}

\maketitle              
\begin{abstract}
Matrix Product State (MPS) methods are among the most effective approaches for the classical simulation of quantum circuits, but their practical performance depends strongly on simulator hyperparameters, and default settings are often suboptimal. In this work, we propose a two-stage framework for automatic hyperparameter selection for quantum MPS simulation. In the first stage, we perform offline single-objective CMA-ES optimization under a fidelity constraint and construct a database of circuit--configuration--performance evaluations. In the second stage, we define a set of static circuit features designed to capture MPS-relevant structural properties and train a circuit-aware hybrid ranking model to recommend configurations for different quantum circuits. We evaluate the approach on multiple scalable circuit families using leave-one-family-out and size-based validation. The results show that offline optimization often improves over default settings, although the magnitude of the gain depends strongly on the backend, circuit family, and circuit scale. The learned predictor recovers a meaningful fraction of this gain, with better performance under size-based validation than under family-based transfer, but generally remains below the offline optimum. 
\keywords{CMA-ES  \and hybrid ranking model \and matrix product states \and quantum circuit simulation \and circuit-aware prediction.}
\end{abstract}
\section{Introduction}

Classical simulation remains a fundamental tool for quantum computing, as current quantum hardware is still constrained by noise and limited scalability~\cite{qc_kim}. It is essential for validating quantum programs and understanding where quantum advantage may realistically emerge~\cite{sim_rev_cicero}. Among classical simulation methods, Matrix Product State (MPS)~\cite{mps_vidal,mps_markov} is one of the most effective practical approaches when the entanglement structure of a circuit remains sufficiently limited, since it can exploit this structure to reduce simulation cost. At the same time, the practical performance of MPS simulation depends strongly on backend hyperparameters~\cite{mps2014}, such as the bond dimension, truncation threshold, measurement strategy, compilation or optimization level, and, in some simulators, additional reordering or contraction controls. These settings determine the trade-off between run time and approximation quality and can strongly affect the feasibility of simulating large circuits. Poor parameter choices may lead either to excessive run time or to loss of accuracy. Identifying an efficient hyperparameter setting is challenging because the search space is noisy, non-linear, and the circuit structure that governs simulation cost is not obvious. The mentioned noise may arise from sources such as floating-point round-off effects and backend-dependent execution behavior. In addition, even small parameter changes can produce large differences in run time and fidelity for large circuits, which makes manual tuning costly and impractical at scale~\cite{bench_leonteva}.

Existing work related to automatic hyperparameter selection for MPS simulation is still limited. Most nearby studies follow broader directions. One focuses on improving MPS simulation for a specific algorithm through a better tensor-network representation and simulation strategy. For Shor's algorithm, earlier work used a static MPS qudit ordering~\cite{mps_wang}, while later work improved the mapping by adapting it to the circuit's entanglement structure and by using measurement to reduce entanglement during sampling~\cite{mps_dang}. Another direction studies optimization of tensor-network methods more generally, through hyper-optimized contraction strategies~\cite{gray_kourtis} or gradient-based optimization of isometric tensor networks~\cite{hauru_riemannian}. Closer to our setting, our previous benchmarking study used single-objective tuning of MPS hyperparameters only to benchmark emulators under near-best conditions~\cite{bench_leonteva}, but its focus was benchmarking rather than optimization or prediction.

In this paper, we address automatic hyperparameter selection for quantum MPS by combining evolutionary optimization with data-driven prediction. We propose a two-stage framework. In the first stage, we run single-objective \texttt{CMA-ES}~\cite{cmaes_hansen} under a fidelity constraint (99\%) and store all evaluated circuit--configuration--performance tuples in a database. This offline stage serves two purposes: it identifies the near-optimal configurations and generates training data by exploring the hyperparameter space more selectively than a uniform sweep. In the second stage, we train a circuit-aware ranking model on this database to predict effective configurations for different quantum circuits from static circuit features. All code used in this study is publicly available in our open-source GitHub repository\footnote{\url{https://github.com/qperfect-io/feniqs_lite}}~\cite{feniqs}. Experiments are performed using the MPS simulators \texttt{Qiskit}~\cite{qiskit} (IBM) and commercial \texttt{MIMIQ}\footnote{\url{https://qperfect.io/mimiq/}} (QPerfect). We evaluate the method on 13 scalable circuit families from the MQTBench\footnote{\url{https://github.com/cda-tum/MQTBench}} library under leave-one-family-out and size-based validation. Our study is therefore guided by three questions. First, how much do offline-tuned hyperparameters improve MPS simulation over default configurations across scalable circuit families? Second, to what extent can this optimized behavior be predicted from circuit structure alone? Third, how do these effects depend on circuit family and scale? By answering these questions, we aim to clarify both the usefulness and the current limits of data-driven hyperparameter selection for quantum MPS simulation.

\section{Proposed Method}

We propose a two-stage framework for automatic hyperparameter selection for Matrix Product State (MPS) simulation of quantum circuits. In the first stage, we perform offline, fidelity-constrained hyperparameter optimization and store the results in a database of circuit--configuration--performance tuples. In the second stage, we use this database to train a circuit-aware predictor that ranks candidate configurations for unseen circuits.

\subsection{Problem formulation}

Let $q$ denote a quantum circuit and let $\boldsymbol{\theta} \in \Theta$ denote a valid MPS simulator configuration. In our setting, $\boldsymbol{\theta}$ is defined by five backend (\texttt{Qiskit}, \texttt{MIMIQ}) hyperparameters presented in Table~\ref{tab:opt_params}. For a given circuit $q$ and configuration $\boldsymbol{\theta}$, the simulator returns a run time $T(q,\boldsymbol{\theta})$ and a fidelity value $F(q,\boldsymbol{\theta})$. Our goal is to identify, for each circuit, a configuration that minimizes run time while satisfying a prescribed fidelity constraint:
\begin{equation}
\boldsymbol{\theta}^{\star}(q)
=
\arg\min_{\boldsymbol{\theta}\in\Theta} T(q,\boldsymbol{\theta})
\quad
\text{s.t.}
\quad
F(q,\boldsymbol{\theta}) \ge \tau ,
\label{eq:constrained_problem}
\end{equation}
where $\tau$ is the minimum acceptable fidelity threshold (99\%).

\subsection{Offline optimization}
\paragraph{Continuous search with discrete simulator parameters.}
We use the Covariance Matrix Adaptation Evolution Strategy (\texttt{CMA-ES})~\cite{cmaes_hansen} because the hyperparameter search problem is derivative-free and noisy. Run time and fidelity measurements may vary across repeated executions, and small changes in value of hyperparameter can lead to highly non-linear changes in performance. Evolutionary strategies are well suited to such settings because they do not rely on gradients and remain effective on irregular objective landscapes. Although \texttt{CMA-ES} is formulated for continuous optimization, we adapt it to discrete parameters by mapping each sampled continuous candidate to the nearest admissible discrete configuration and using a margin correction of the covariance matrix to avoid premature convergence~\cite{cmaes_discrete}. 
So, we optimize over a continuous vector
\[
\mathbf{x} = (x_1,\dots,x_d) \in [0,1]^d,
\]
where $d$ is the number of tunable hyperparameters (5 for both selected backends).

For each hyperparameter $j$, let $m_j$ be the number of admissible discrete values for that parameter. We map the continuous coordinate $x_j$ to a discrete index by
\begin{equation}
k_j =
\mathrm{clip}\!\left(
\mathrm{round}\!\left(x_j (m_j - 1)\right),
0,
m_j - 1
\right).
\label{eq:index_mapping}
\end{equation}
The selected index $k_j$ is then used to choose the corresponding admissible value of parameter $j$. Applying this rule to all coordinates defines a deterministic projection from the continuous search space to a valid simulator configuration, denoted by
\[
\phi(\mathbf{x}) \in \Theta.
\]

\paragraph{Fidelity-constrained objective.}

For each sampled vector $\mathbf{x}$, we evaluate the discretized configuration $\phi(\mathbf{x})$ several times and compute the average run time and average fidelity,
\[
\overline{T}(q,\phi(\mathbf{x})) \quad \text{and} \quad \overline{F}(q,\phi(\mathbf{x})).
\]

The optimization is single-objective: we minimize run time while enforcing a fidelity threshold $\tau$. The scalar objective optimized by \texttt{CMA-ES} is defined as
\begin{equation}
J(q,\mathbf{x}) =
\begin{cases}
\overline{T}(q,\phi(\mathbf{x})),
& \text{if } \overline{F}(q,\phi(\mathbf{x})) \ge \tau, \\[6pt]
M + \lambda \bigl(\tau - \overline{F}(q,\phi(\mathbf{x}))\bigr),
& \text{if } 0 \le \overline{F}(q,\phi(\mathbf{x})) < \tau, \\[6pt]
M_{\mathrm{fail}},
& \text{if the simulator execution fails,}
\end{cases}
\label{eq:penalized_objective}
\end{equation}
where $M$ is a large penalty constant, $\lambda > 0$ controls the severity of fidelity violations, and $M_{\mathrm{fail}}$ is a very large penalty assigned to failed executions.

\paragraph{Margin-aware covariance correction.}

Because discretization may cause many continuous samples to collapse onto the same discrete configuration, \texttt{CMA-ES} can lose diversity prematurely. To mitigate this effect, we use a margin-aware covariance correction. For each sampled vector, the optimizer records the distance between the scaled continuous coordinate and the selected discrete index. If the probability of crossing a discrete boundary becomes too small, the covariance in that coordinate is increased. This preserves exploration near quantization boundaries and improves search robustness in the mixed continuous--discrete setting.

\paragraph{Repeated evaluations and caching}

To reduce noise, each discretized configuration is executed multiple times, and the average run time and average fidelity are used in~\eqref{eq:penalized_objective}. Since different continuous vectors may map to the same discrete configuration, evaluations are cached using the discretized parameter tuple as a key. If a configuration has already been evaluated, the cached result is reused instead of running the simulator again. This substantially reduces the number of backend calls and improves the efficiency of offline optimization.

Running the above procedure over a collection of training circuits produces a database
\[
\mathcal{D} = \{(q_i, \boldsymbol{\theta}_j, T_{ij}, F_{ij})\},
\]
where each tuple contains a circuit, a valid MPS configuration, and its measured run time and fidelity.

\subsection{Online prediction}

The online stage is formulated as a ranking problem~\cite{liu_rank} over candidate simulator configurations. For each circuit \(q\), the offline optimization stage provides a set of evaluated configurations \(\boldsymbol{\theta}\in\Theta\), together with their measured run time \(T(q,\boldsymbol{\theta})\) and fidelity \(F(q,\boldsymbol{\theta})\). The predictor learns from these data and, for a new circuit, ranks the candidate configurations in order to recommend the most promising one.

\paragraph{Learning targets.}

For each circuit \(q\), let
\[
F^{\star}(q)=\max_{\boldsymbol{\theta}\in\Theta}F(q,\boldsymbol{\theta})
\]
denote the best fidelity observed in the offline database. Given a small tolerance \(\varepsilon>0\), we define the best feasible run time as
\begin{equation}
T^{\star}_{\mathrm{feas}}(q)
=
\min_{\boldsymbol{\theta}\,:\,F(q,\boldsymbol{\theta})\ge F^{\star}(q)-\varepsilon}
T(q,\boldsymbol{\theta}),
\label{eq:best_feasible_runtime_online}
\end{equation}
that is, the smallest run time among configurations whose fidelity is sufficiently close to the best observed one.

For every circuit--configuration pair \((q,\boldsymbol{\theta})\), we compute the run time ratio
\begin{equation}
r(q,\boldsymbol{\theta})
=
\frac{T(q,\boldsymbol{\theta})}{T^{\star}_{\mathrm{feas}}(q)},
\label{eq:runtime_ratio_online}
\end{equation}
where \(r=1\) corresponds to the best feasible run time, and the fidelity gap
\begin{equation}
\Delta_F(q,\boldsymbol{\theta})
=
\max\!\bigl(0,\;F^{\star}(q)-F(q,\boldsymbol{\theta})\bigr),
\label{eq:fidelity_gap_online}
\end{equation}
which measures how far the fidelity is below the best value observed for this circuit.

A configuration is labeled as near-optimal if it is close to the best feasible configuration in both run time and fidelity:
\begin{equation}
\mathrm{NearOpt}(q,\boldsymbol{\theta})=1
\iff
r(q,\boldsymbol{\theta})\le 1.05
\;\land\;
\Delta_F(q,\boldsymbol{\theta})\le \varepsilon.
\label{eq:near_opt_online}
\end{equation}

In addition, we define a continuous target score
\begin{equation}
s(q,\boldsymbol{\theta})
=
-\left[
\log\!\bigl(\max(r(q,\boldsymbol{\theta}),1)\bigr)
+
\alpha\,\Delta_F(q,\boldsymbol{\theta})
\right],
\label{eq:target_score_online}
\end{equation}
where \(\alpha>0\) controls how strongly fidelity degradation is penalized. Higher values of \(s(q,\boldsymbol{\theta})\) correspond to better configurations. In the implementation, we also learn two auxiliary regression targets: the run-time penalty \(\log(\max(r(q,\boldsymbol{\theta}),1))\) and the fidelity penalty \(\Delta_F(q,\boldsymbol{\theta})\).

\paragraph{Circuit features.}
The predictor uses a list of 107 features extracted from the OpenQASM description of the circuit. These features are designed to capture properties that are relevant for MPS simulation cost. We group them into 6 categories according to the aspect of circuit structure they describe: global circuit scale, gate composition, interaction graph, and locality/temporal structure.

The first group describes the overall scale of the circuit: the number of qubits, the number of classical bits, the total number of operations, the circuit depth, and normalized depth measures such as depth per qubit. These quantities distinguish small and shallow circuits from larger and more computationally dense ones. 

The second group describes gate composition. It includes counts and fractions of major gate categories, such as one-qubit gates, two-qubit gates, entangling gates, controlled gates, phase gates, rotation gates, measurements. This is important because MPS cost is sensitive to the amount and arrangement of entangling operations. 

The third group summarizes the interaction graph induced by multi-qubit gates. In this graph, qubits are vertices and an edge is added whenever two qubits interact through a two-qubit gate. From this graph, we compute: the number of distinct interacting pairs; the interaction density, that is, the fraction of possible qubit pairs that actually interact; edge reuse, that is, how often the same qubit pair is used repeatedly; degree statistics, which describe how unevenly interactions are distributed across qubits; span statistics, which measure how far apart interacting qubits are in the circuit ordering; and entropy-based quantities, which measure how concentrated or how diffuse the interaction pattern is. These descriptors help distinguish localized and repetitive interaction patterns from broad and irregular ones. 

The fourth group captures locality and temporal structure. These features include the fraction of nearest-neighbor versus long-range interactions, cut-based proxies that estimate how much two-qubit activity crosses a bipartition of the qubit line, the amount of entangling activity appearing late in the circuit, burstiness of gate activity over depth, and simple regularity measures of the gate sequence. Such features are especially relevant for MPS simulation, since long-range couplings and irregular surges of entangling gates tend to increase intermediate bond dimensions and make truncation more difficult. 

Finally, we include summary statistics of gate angles and phase parameters, such as the diversity of parameter values, their numerical spread, sign balance, and the fraction of small or highly structured angles. These descriptors help distinguish circuits with similar gate counts but different numerical behavior under truncation.

\paragraph{Candidate representation.} Each candidate configuration is represented not only by its raw hyperparameter values, but also by transformed descriptors that better reflect how these parameters affect MPS simulation. In particular, the bond-dimension parameter is encoded on a logarithmic base-2 scale because it takes values that grow by powers of two, and its effect on memory and computational cost is more naturally related to relative doubling than to absolute linear increments. Similarly, the precision- or truncation-related parameter is encoded on a logarithmic base-10 scale because it spans several orders of magnitude, so differences are meaningful primarily at the level of decimal orders rather than raw values. In addition, we include interaction terms to capture nonlinear dependencies between circuit stress, approximation strength, optimization level, and backend-specific options.


\paragraph{Hybrid ranking model.}
The ranking model itself is a multi-objective hybrid ensemble. For each circuit--candidate pair \((q,\boldsymbol{\theta})\), the model combines: (i) three regressors predict the main score in Equation~\eqref{eq:target_score_online}; (ii) two regressors predict the run-time penalty; (iii) two regressors predict the fidelity penalty; (iv) and one classifier predicts the probability that the candidate is near-optimal. The regression components are based on tree-ensemble methods, namely random forests~\cite{breiman_rf}, extremely randomized trees~\cite{geurts_erf}, and gradient-boosted decision trees, while the classification component is based on extremely randomized trees~\cite{geurts_erf}. These models were selected because they can capture non-linear relationships and feature interactions and remain robust on structured data without requiring large training sets.


Let \(\hat{s}_1(q,\boldsymbol{\theta})\), \(\hat{s}_2(q,\boldsymbol{\theta})\), and \(\hat{s}_3(q,\boldsymbol{\theta})\) denote the outputs of the three primary regressors for the circuit--candidate pair \((q,\boldsymbol{\theta})\). Here, the hat symbol indicates a model prediction. We define their average prediction as
\begin{equation}
\bar{s}_{\mathrm{prim}}(q,\boldsymbol{\theta})
=
\frac{1}{3}\sum_{i=1}^{3}\hat{s}_i(q,\boldsymbol{\theta}),
\label{eq:primary_average}
\end{equation}
Let \(\hat{r}_{\mathrm{pen}}(q,\boldsymbol{\theta})\) denote the predicted run-time penalty, \(\hat{f}_{\mathrm{pen}}(q,\boldsymbol{\theta})\) the predicted fidelity penalty, and \(\hat{p}(q,\boldsymbol{\theta})\) the predicted probability that the candidate belongs to the near-optimal class. To quantify the disagreement between the three primary regressors, we also define the uncertainty estimate
\begin{equation}
u(q,\boldsymbol{\theta})
=
\sqrt{
\frac{1}{3}\sum_{i=1}^{3}
\left(
\hat{s}_i(q,\boldsymbol{\theta})
-
\bar{s}_{\mathrm{prim}}(q,\boldsymbol{\theta})
\right)^2
},
\label{eq:primary_uncertainty}
\end{equation}
that is, the standard deviation of the three primary predictions.

The final ranking score is then defined as
\begin{equation}
\hat{s}(q,\boldsymbol{\theta})
=
\bar{s}_{\mathrm{prim}}(q,\boldsymbol{\theta})
+
\beta\!\left(\hat{p}(q,\boldsymbol{\theta})-\frac{1}{2}\right)
-
\gamma\,\hat{r}_{\mathrm{pen}}(q,\boldsymbol{\theta})
-
\delta\,\hat{f}_{\mathrm{pen}}(q,\boldsymbol{\theta})
-
\eta\,u(q,\boldsymbol{\theta}),
\label{eq:hybrid_score_online}
\end{equation}
where \(\beta,\gamma,\delta,\eta\) are fixed mixing coefficients for balancing the contributions of the averaged primary score, set to \(\beta=0.35\), \(\gamma=0.40\), \(\delta=6.0\), and \(\eta=0.10\), and kept unchanged for all experiments. They were chosen empirically to balance score components: e.g., \(\delta\) is larger because fidelity-gap predictions are numerically smaller, while \(\eta\) remains small since the uncertainty term is used only as a secondary correction. Thus, the first term rewards candidates with high predicted quality, the second provides a confidence-based correction, the third and fourth penalize expected run-time and fidelity losses, and the last penalizes disagreement between the primary regressors. The classifier contribution is introduced through the centered term \(\beta\!\left(\hat{p}(q,\boldsymbol{\theta})-\frac{1}{2}\right)\), so that it acts as a positive correction when the predicted probability of belonging to the near-optimal class exceeds \(0.5\), as a penalty when it is below \(0.5\), and has no effect when the classifier is maximally uncertain.  For a new circuit, all candidates are ranked in decreasing order of \(\hat{s}(q,\boldsymbol{\theta})\), and the top-ranked configuration is returned as the recommended MPS setting.
\section{Results}

\subsection{Experimental Setup}
All experiments were performed within our open-source benchmarking framework, \texttt{FENIQS}\footnote{\url{https://github.com/qperfect-io/feniqs_lite}}, which provides the full pipeline, including hyperparameter optimization, feature extraction, model training, and validation. As benchmark suite, we use the same scalable \texttt{OpenQASM~2.0} circuits as in our earlier benchmarking study on utility-scale quantum emulators~\cite{bench_leonteva}, in order to keep the evaluation setting consistent. The suite consists of 13 circuit families~\cite{bench_leonteva}, spanning standard algorithmic primitives and application-oriented workloads, including amplitude estimation~\cite{ae}, quantum Fourier transform~\cite{qft}, exact and inexact phase estimation~\cite{qpeexact}, entangled-state preparation~\cite{ghz},~\cite{wstate}, variational ans\"atze~\cite{realamp}, quantum walks~\cite{qwalk}, random circuits~\cite{random}, and quantum neural networks~\cite{qnn}. These circuits originate from the open-source \texttt{MQTBench} library\footnote{\url{https://github.com/cda-tum/MQTBench}}~\cite{mqt_bench}, remain scalable beyond 100 qubits, and were transpiled to a minimal gate set, sanitized for cross-backend compatibility, and released together with their mirror-circuit versions on Zenodo\footnote{\url{https://doi.org/10.5281/zenodo.15220683}}.

In our experiments, the benchmark suite is not uniformly distributed across qubit sizes. This is because some families, especially those with more complex entangling structure, become prohibitively expensive for MPS simulation already at moderate sizes; for example, for random circuits we could evaluate only instances up to about 26--28 qubits~\cite{bench_leonteva}. Table~\ref{tab:family_instances} summarizes, for each circuit family, the number of instances, where each instance corresponds to a circuit at a given qubit size, together with the largest qubit size included in the study. To prevent families with more training instances from having a disproportionate influence on the learned model, we weight each training sample inversely to the size of its circuit family.

\begin{table}
\centering
\caption{Benchmark coverage across circuit families in terms of instances and maximum qubit size.}
\label{tab:family_instances}
\resizebox{\textwidth}{!}{%
\begin{tabular}{lccccccccccccc}
\hline
Family & \texttt{ae} & \texttt{ghz} & \texttt{graphstate} & \texttt{qft} & \texttt{qftent} & \texttt{qnn} & \texttt{qpeexact} & \texttt{qpeinexect} & \texttt{qwalk} & \texttt{random} & \texttt{realamprandom} & \texttt{su2random} & \texttt{wstate} \\
\hline
\# instances & 24 & 34 & 18 & 34 & 34 & 12 & 34 & 34 & 12 & 8 & 10 & 10 & 36 \\
max nb. qubits & 100 & 832 & 44 &240 & 240 & 24 & 240 & 240 & 24 & 12 & 16 & 16 & 1024 \\
\hline
\end{tabular}%
}
\end{table}

We focus on the MPS backends of \texttt{Qiskit} (open-source by IBM) and \texttt{MIMIQ} (commercial by QPerfect), since our earlier benchmarking study identified these emulators as the most effective overall on the considered scalable benchmark suite~\cite{bench_leonteva}. Accuracy of MPS simulation is evaluated using \emph{mirror-circuit fidelity}~\cite{proctor_mirfid}, a backend-independent fidelity estimate that provides a common accuracy measure across different simulators. In this approach, a circuit is followed by its reverse, and fidelity is measured as the probability of returning to the initial state. Run time is defined as the total wall-clock time
required to execute a quantum circuit (excluding mirror
circuit fidelity estimation)~\cite{bench_leonteva}.

We evaluate the proposed approach in two parts. First, we assess the quality of the hyperparameter configurations obtained by \texttt{CMA-ES} optimization. Second, we evaluate the predictive model through cross-validation on the optimization-derived dataset, including transfer across circuit families and across qubit sizes.

\subsection{Validation of Offline \texttt{CMA-ES} Optimization}
For \texttt{Qiskit} and \texttt{MIMIQ}, we optimize five discrete hyperparameters each: the corresponding search spaces are summarized in Table~\ref{tab:opt_params}. For both backends, the meaning and practical role of the optimized parameters are described in the official simulator documentation: for \texttt{Qiskit}\footnote{\url{https://qiskit.github.io/qiskit-aer/stubs/qiskit_aer.AerSimulator.html}} and for \texttt{MIMIQ}\footnote{\url{https://docs.qperfect.io/MimiqCircuits.jl/stable/manual/simulation/\#Matrix-Product-States}}.
In all experiments, \texttt{CMA-ES} uses a population size of 10 candidate solutions and is run for 15 iterations.
Optimization performance is reported as the speedup of the default configuration relative to the optimized one, so values greater than one indicate an improvement in run time. Each candidate configuration is evaluated in the same execution environment and with the same \emph{mirror-fidelity} protocol.

\begin{table}
\centering
\caption{Optimized hyperparameters and their admissible values for the two MPS backends.}
\label{tab:opt_params}
\begin{tabular}{lll}
\hline
Backend & Hyperparameter & Admissible values \\
\hline
\multirow{5}{*}{\texttt{Qiskit}}
& Maximum bond dimension
& \{4, 8, 16, 32, 64, 128, 256, 512, 1024, 2048\} \\
& LAPACK flag
& \{0, 1\} \\
& Compilation level
& \{1, 2, 3\} \\
& Truncation threshold
& \(\{10^{-1}, 10^{-2}, 10^{-3}, 10^{-4}, 10^{-5}, 10^{-6}, 10^{-7}, 10^{-8}\}\) \\
& Measurement strategy
& \{\texttt{mps\_probabilities}, \texttt{mps\_apply\_measure}\} \\
\hline
\multirow{5}{*}{\texttt{MIMIQ}}
& Bond dimension
& \{4, 8, 16, 32, 64, 128, 256, 512\} \\
& Entanglement dimension
& \{4, 8, 16\} \\
& Optimization level
& \{0, 1\} \\
& Truncation cutoff
& \(\{10^{-1}, 10^{-2}, 10^{-3}, 10^{-4}, 10^{-5}, 10^{-6}, 10^{-7}, 10^{-8}\}\) \\
& Simulation method
& \{\texttt{vmpoa}, \texttt{dmpo}\} \\
\hline
\end{tabular}
\end{table}

\begin{figure}
    \centering
    \includegraphics[width=\textwidth]{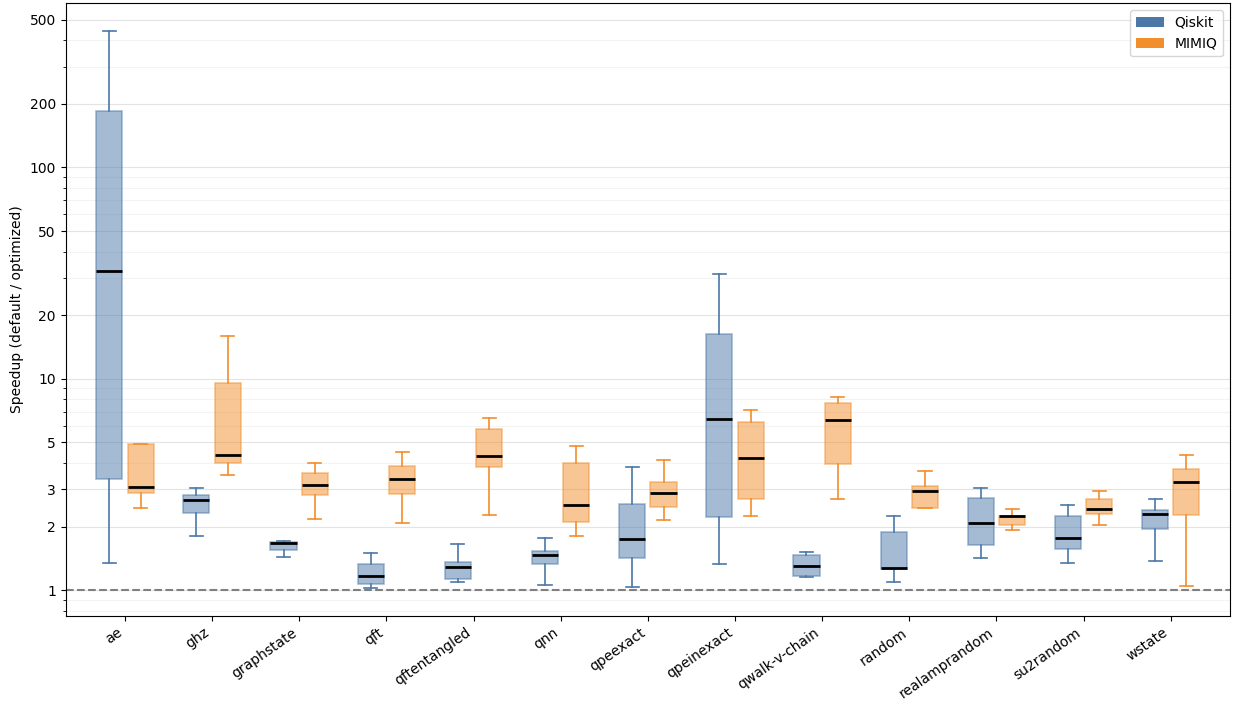}
    \caption{Speedup (ratio of default params. run time / optimized param. run time) by circuit family for \texttt{Qiskit} (blue) and \texttt{MIMIQ} (orange). Each box plot represents the distribution of speedup values across circuit instances with varying qubit counts. The median is shown by the central line, and the dashed horizontal line at 1 indicates no improvement. The vertical axis is shown on a logarithmic scale.}
    \label{fig:cmaes_vs_default}
\end{figure}
Figure~\ref{fig:cmaes_vs_default} compares the distribution of offline optimization speedup across circuit families for \texttt{Qiskit} and \texttt{MIMIQ}. Speedup is defined as the ratio of default to optimized run time, so values above one indicate improvement. For each circuit family, the box plot aggregates the speedup values obtained for the available circuit instances at different qubit counts. The central line indicates the median, the box spans the interquartile range, and the whiskers show the spread of the observed values. Compact distributions indicate more consistent gains across instances, whereas wider ones indicate larger variability with circuit size.

The two backends exhibit clearly different behavior. For \texttt{Qiskit}, strong gains are concentrated in a small number of families: \texttt{ae} is by far the most favorable case, with a median speedup of about \(32\times\), followed by \texttt{qpeinexact} with about \(6.4\times\). A second group, including \texttt{ghz}, \texttt{wstate}, and \texttt{realamprandom}, reaches median speedups around \(2\times\) to \(3\times\), while the remaining families stay much closer to the default configuration, typically around \(1.2\times\) to \(1.8\times\). By contrast, \texttt{MIMIQ} shows a broader optimization effect across families. Its largest median gains are observed for \texttt{qwalk-v-chain} (\(\approx 6.4\times\)), \texttt{ghz} (\(\approx 4.4\times\)), \texttt{qftentangled} (\(\approx 4.3\times\)), and \texttt{qpeinexact} (\(\approx 4.2\times\)), while several other families, including \texttt{qft}, \texttt{wstate}, \texttt{graphstate}, \texttt{random}, \texttt{ae}, and \texttt{qpeexact}, still show median gains around \(3\times\).

Overall, these results show that offline hyperparameter optimization is beneficial for both backends, but its magnitude depends on both circuit family and qubit count. For \texttt{Qiskit}, the strongest improvements are concentrated in a few families and become more pronounced at larger sizes, whereas for \texttt{MIMIQ} the optimization effect is typically more evenly distributed across the available size range.

\subsection{Validation of Online Prediction}
The predictive model is trained on the optimization database constructed during the offline CMA-ES stage. Each data point consists of a circuit description, extracted features, and the corresponding optimized hyperparameter configuration. This ensures that the model learns from configurations evaluated under the same execution environment and fidelity constraint as used in the optimization stage. 

To assess the generalization capability of the model, we consider two complementary validation settings: (i) Family-based validation -- in the leave-one-family-out setting, all circuits belonging to one family are held out for testing, while the model is trained on all remaining families. This evaluates whether the predictor can transfer to previously unseen circuit types; (ii) Size-based validation -- in the size-based setting, validation is performed separately within each circuit family. For a given family, we hold out all circuits corresponding to one qubit size, train the model on the remaining qubit sizes of the same family, and then evaluate on the held-out size. This procedure is repeated over the available qubit sizes, so that each selected size is used as a test case in turn. It evaluates whether the model can extrapolate across scale, that is, whether it can recommend effective hyperparameters for unseen system sizes after learning from other sizes within the same circuit class.

\begin{figure}
    \centering
    \includegraphics[width=\textwidth]{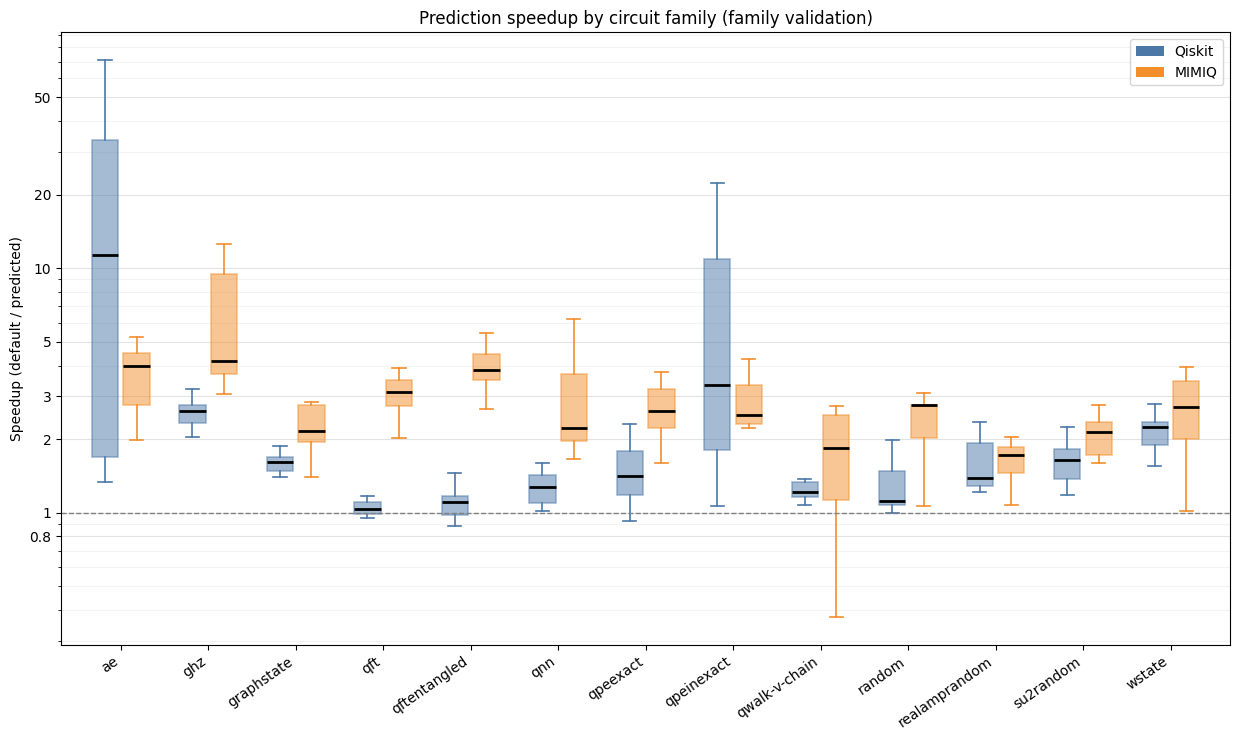}
    \caption{Prediction speedup (ratio of default to predicted run time) by circuit family for \texttt{Qiskit} (blue) and \texttt{MIMIQ} (orange) under one-family-out validation. The black line shows the median, boxes the interquartile range, and whiskers the variability across instances. The dashed line at 1 indicates no improvement. The vertical axis is logarithmic.
    }
    \label{fig:pred_vs_default}
\end{figure}

\begin{figure}
    \centering
    \includegraphics[width=\textwidth]{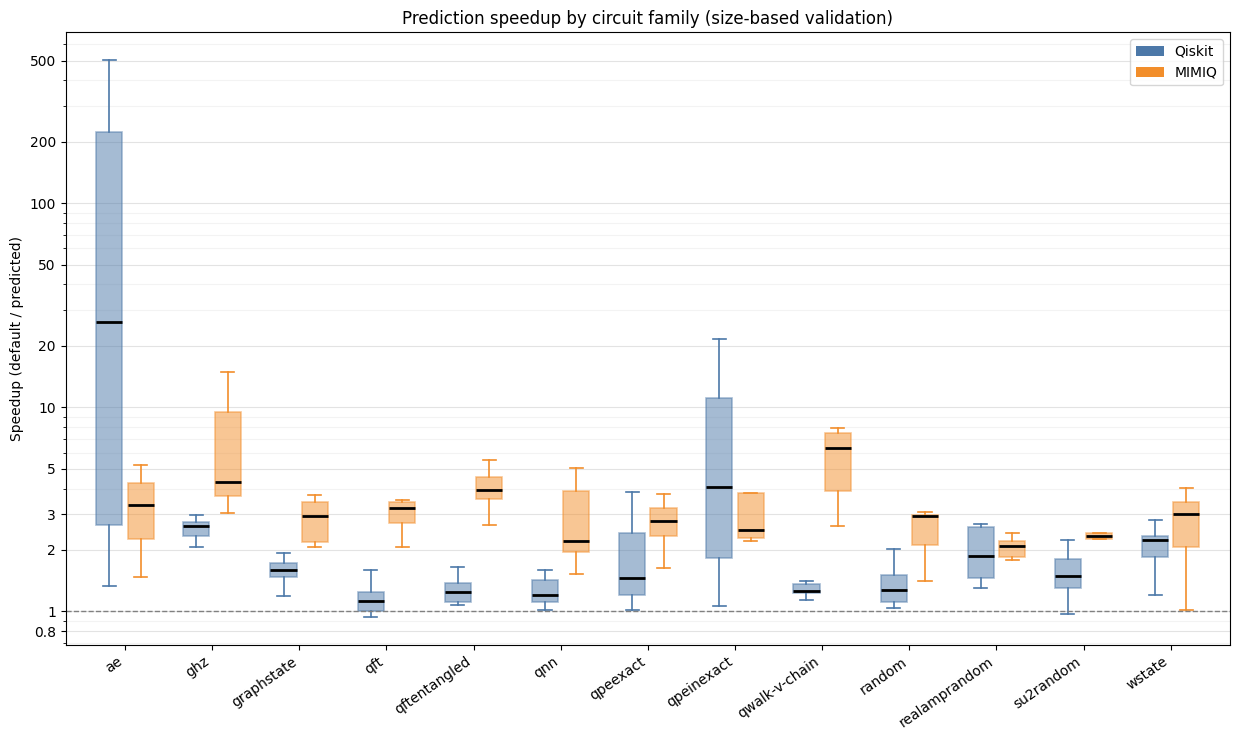}
    \caption{Prediction speedup (ratio of default to predicted run time) by circuit family for \texttt{Qiskit} (blue) and \texttt{MIMIQ} (orange) under size-based validation. The black line shows the median, boxes the interquartile range, and whiskers the variability across instances. The dashed line at 1 indicates no improvement. The vertical axis is shown on a logarithmic scale.}
    \label{fig:pred_vs_default_by_size}
\end{figure}

We evaluate the effectiveness of the predicted hyperparameters by comparing their performance against the simulator default configuration. For each circuit instance, speedup is defined as the ratio of execution time using default parameters to execution time using predicted parameters. Figures~\ref{fig:pred_vs_default} and~\ref{fig:pred_vs_default_by_size} report the distribution of speedup across circuit families for family-based and size-based validation, respectively.

 The results are better under size-based validation for both backends, although the gain from this change is much more pronounced for \texttt{Qiskit} than for \texttt{MIMIQ}. For \texttt{Qiskit}, the family-based setting already recovers the main offline pattern: the largest median gains are obtained for \texttt{ae} (\(\approx 11.3\times\)) and \texttt{qpeinexact} (\(\approx 3.3\times\)), followed by \texttt{ghz} and \texttt{wstate} (\(\approx 2.6\times\) and \(\approx 2.2\times\)). All remaining families stay much closer to the default setting, typically between about \(1.0\times\) and \(1.7\times\). Under size-based validation, the same ranking is preserved, but the gains increase, most notably for \texttt{ae}, whose median rises to about \(26.1\times\), and for \texttt{qpeinexact}, which reaches about \(4.1\times\). At the same time, the fraction of instances below the default baseline decreases from about \(8.4\%\) in the family-based setting to about \(3.5\%\) in the size-based setting. This indicates that for \texttt{Qiskit} it is easier to extrapolate across circuit size within a known family than to transfer to a completely unseen family. This is consistent with the offline results, where the strongest \texttt{Qiskit} gains were concentrated in the same families and became more pronounced at larger qubit counts. For \texttt{MIMIQ}, both family-based and size-based validation recover a substantial fraction of the offline optimization gain. In most circuit families, the predicted median speedup remains below the optimized reference, but the gap is often moderate. For example, \texttt{qftentangled} decreases from about \(4.30\times\) offline to \(3.83\times\) under family-based and \(3.94\times\) under size-based validation, while \texttt{qpeinexact} drops more noticeably from about \(4.23\times\) to \(2.50\times\) in both settings. At the same time, several families remain very close to the offline result, including \texttt{ghz} (\(4.38\times\) offline versus \(4.16\times\) under family-based and \(4.32\times\) under size-based validation), \texttt{random} (\(2.94\times\) offline versus \(2.76\times\) and \(2.94\times\)), and \texttt{qwalk-v-chain} (\(6.36\times\) offline versus \(6.30\times\) under size-based, but \(1.85\times\) under family-based validation). The only family for which prediction exceeds the current offline reference is \texttt{ae}, whose median speedup increases from about \(3.08\times\) offline to \(3.99\times\) under family-based and \(3.32\times\) under size-based validation. The fact that prediction can exceed the recorded offline result is possible because the \texttt{CMA-ES} search is itself approximate and performed under a limited budget (10 solutions $\times$ 15 iterations), so the selected offline configuration is not guaranteed to be globally optimal. However, for \texttt{Qiskit}, prediction does not exceed the offline optimized reference at the family-median level. Overall, prediction against the default setting is useful for both backends, but its success depends on both circuit family and validation regime, with size-based extrapolation being consistently easier than family transfer.



\section{Conclusion}
Our study was guided by three questions: how much offline-tuned hyperparameters improve MPS simulation over default settings, how the results depend on circuit family and circuit scale, and to what extent this optimized behavior can be predicted from circuit structure. The results allow us to answer all three questions, which we summarize below.

Offline hyperparameter tuning is often beneficial for both \texttt{Qiskit} and \texttt{MIMIQ}, but to a different extent. For \texttt{Qiskit}, large gains are concentrated in only a few families (and become more pronounced at larger qubit numbers), most notably \texttt{ae} and \texttt{qpeinexact}, while many others remain relatively close to the default configuration. For \texttt{MIMIQ}, the gains are more broadly distributed across the benchmark suite, with several families showing median improvements of roughly \(3\times\) to \(6\times\). Thus, offline tuning is useful for both backends, but its effect is strongly backend- and family-dependent; both circuit family and circuit scale matter. 

The differences in optimization results between \texttt{Qiskit} and \texttt{MIMIQ} can likely be explained by the way the two backends control approximation, since in MPS simulation, computational cost depends not only on the entangling structure of the circuit, but also on how the backend regulates the approximation regime during execution.

The optimized behavior is only partially predictable from circuit structure alone. In both backends, the predictor often recovers a substantial fraction of the offline gain, but not all of it. The results are generally better under size-based validation than under leave-one-family-out validation, which indicates that once the model has seen circuits from a family, it can extrapolate across qubit counts more reliably than it can transfer to a completely unseen family. Prediction is therefore practically useful, but it does not replace direct optimization.

The benchmark set is not statistically uniform across circuit sizes in the present study; however, despite this limitation, the comparison between validation regimes remains informative and can be refined further in future work using a larger and more balanced dataset.

Beyond hyperparameter selection itself, we believe that the circuit features introduced in this work may also be useful for broader quantum-circuit optimization tasks. Since they capture structural properties such as circuit scale, gate composition, interaction patterns, locality, and temporal organization, they could help identify suitable optimization strategies in compilation routines for a given circuit.

Overall, the study shows that data-driven hyperparameter selection for MPS simulation is both useful and limited. Offline optimization provides a clear improvement over default settings, and learning can transfer part of this gain to unseen circuits. At the same time, the remaining gap between prediction and offline tuning shows that broader training data is still needed.

\subsubsection{\discintname}
Author G.Masella is a shareholder of QPerfect. 
%
%
%
%

\end{document}